\documentclass[aps,pre,floatfix,superscriptaddress,twocolumn,showkeys,10pt]{revtex4-1}
\usepackage{color}
\usepackage{graphicx}
\usepackage{amsmath, amsthm, amssymb}
\usepackage[colorlinks,citecolor=red,urlcolor=blue]{hyperref}
\newcommand{\be}{\begin{equation}}
\newcommand{\ee}{\end{equation}}
\newcommand{\bea}{\begin{eqnarray}}
\newcommand{\eea}{\end{eqnarray}}
\usepackage{graphicx}
\usepackage{epstopdf}
\usepackage{color}
\definecolor{orange}{RGB}{255,127,0}
\definecolor{blue2}{RGB}{33,114,173}

\begin{document}
\title{Aging of ring polymeric topological glass formers via thermal quench}

\author{Arabinda Behera}
\email{arabindab@imsc.res.in}
\affiliation{The Institute of Mathematical Sciences, C.I.T. Campus,
Taramani, Chennai 600113, India}
\affiliation{Homi Bhabha National Institute, Training School Complex, Anushakti Nagar, Mumbai, 400094, India}

\author{Projesh Roy}
\email{projesh@nitt.edu}
\affiliation{Department of Chemistry, National Institute of Technology Tiruchirappalli, Tiruchirappalli 620015, India}

\author{Pinaki Chaudhuri}
\email{pinakic@imsc.res.in}
\affiliation{The Institute of Mathematical Sciences, C.I.T. Campus,
Taramani, Chennai 600113, India}
\affiliation{Homi Bhabha National Institute, Training School Complex, Anushakti Nagar, Mumbai, 400094, India}

\author{Satyavani Vemparala}
\email{vani@imsc.res.in}
\affiliation{The Institute of Mathematical Sciences, C.I.T. Campus,
Taramani, Chennai 600113, India}
\affiliation{Homi Bhabha National Institute, Training School Complex, Anushakti Nagar, Mumbai, 400094, India}
\date{\today}
\begin{abstract}
We investigate the dynamical response of glass-forming systems composed of topologically constrained ring polymers subjected to an instantaneous thermal quench, employing large-scale molecular dynamics simulations. We demonstrate that the onset of glassiness depends on polymer stiffness, with increased rigidity enhancing configurational constraints and delaying structural relaxation. In the glassy regime, the system exhibits hallmark aging characteristics, as evidenced by two-time correlation functions, namely the mean square displacement and self-intermediate scattering function, which display a clear dependence on the waiting time following the thermal quench. The extracted relaxation timescale ($\tau_\alpha$) follows an approximate simple aging scenario with waiting time ($t_w$), described by $\tau_\alpha \sim t_w^b$, where $0.8 < b < 0.93$. Finally, we analyze the threading of rings during the thermal quench, demonstrating that both increased and persistent threading correlate with the emergence of glassiness. Moreover, the threading persistence timescale exhibits a strong correlation with the structural relaxation timescale. Our study thus provides a comprehensive view of structural relaxation and aging in dense ring polymer systems, highlighting the critical roles of topological constraints and polymer stiffness in governing non-equilibrium glassy dynamics.
\end{abstract}

\keywords{Ring polymers, topological glass, glass transition, aging, molecular dynamics}

\maketitle

\section{\label{sec:intro} Introduction}
Glass-forming liquids transform into a glassy state when rapidly quenched to low temperatures~\cite{binder2011glassy}. Empirically, the glass transition temperature is defined as the temperature at which the material becomes exceedingly viscous, typically around $10^{12}$ poise~\cite{angell1995formation}. Below this transition temperature, the material struggles to reach equilibrium within typical observation timescales, and its properties continue to temporally evolve  even when held at a constant temperature—this phenomenon is known as aging~\cite{bouchaud2000aging,bandyopadhyay2006slow,arceri2022glasses,janzen2022aging}. Such out-of-equilibrium behavior arises due to the complex underlying energy landscape of glass-forming materials, where the system becomes trapped in local minima, resulting in extremely slow relaxation dynamics~\cite{debenedetti2001theory, bouchaud2000aging}. Aging behavior has been observed in various glassy materials, including colloidal glasses, polymers, and metallic glasses, where structural and dynamical properties exhibit slow temporal evolution~\cite{pusey1987observation,yang2022configurational,lu2000correlation,dudowicz2005glass,jin2022molecular,mckenna2003mechanical,hufnagel2016deformation}. Understanding aging remains a central question in exploring the slow dynamics and long-term stability of these materials.

Ring polymer systems with varying stiffness serve as excellent proxies for studying the dynamics of colloidal particles, particularly regarding glass formation and aging in soft, deformable systems. Due to their closed-loop architecture, ring polymers exhibit unique rheological properties distinct from their linear counterparts. Several studies have demonstrated that ring polymers display unconventional stress-relaxation behavior, primarily due to the absence of free ends. This prevents reptation-like motion and instead creates entanglement constraints mediated by threading effects~\cite{Rosa_Everaers_PhysRevLett_2014, Halverson_Kremer_PhysRevLett_2012, Halverson_Kremer_JChemPhys_2011, Halverson_Kremer_JChemPhys_2011_2, rubinstein1986dynamics, cates1986conjectures, lee2015segregated, brown2001influence, suzuki2009dimension, Lee_Jung_Polymers_2019, deguchi2017statistical}. Such topological constraints result in dynamical slowdowns analogous to crowding effects observed in dense colloidal suspensions, making ring polymers valuable models for exploring glassy dynamics. By tuning the backbone stiffness of ring polymers, their phase behavior can be systematically controlled, akin to adjusting interactions in soft colloidal systems to modulate their proximity to the glass transition~\cite{mukhopadhyay2011packings, makse2000packing, batista2010crystallization, boromand2018jamming, tu2023unexpected}.

Despite these similarities, the aging dynamics of topologically constrained ring polymers remain underexplored. While studies of linear polymers have shown that chain stiffness and density significantly influence the glass transition temperature~\cite{privalko1974glass, dudowicz2005glass, hsu2019coarse, ness2017nonmonotonic}, ring polymers add an additional layer of complexity due to threading interactions. Unlike traditional colloidal or polymeric glasses, where relaxation is primarily governed by caging or linear-chain entanglements, ring polymers experience unique topological frustrations arising from threading-mediated constraints, significantly hindering structural relaxation and reorganization~\cite{Obukhov_Duke_PhysRevLett_1994, Obukov_Colby_Macromol_1994, Guo_Zhang_Polymers_2020, stavno2023thread, lee2015slowing, roy2022effect}. Previous studies have indicated that chain stiffness influences glassy dynamics in ring polymer systems with mixed stiffnesses~\cite{roy2022effect, roy2024bidisperse}; however, a systematic characterization of aging dynamics as a function of stiffness remains lacking. Investigating how polymer stiffness affects threading dynamics and determining the glass transition temperature of ring polymer systems could provide critical insights into how topological constraints influence aging and structural relaxation, deepening our understanding of glass formation in these unique polymeric systems.

\section{\label{sec:methods}Methods}

\begin{figure}
\centering
 \includegraphics[width=\columnwidth]{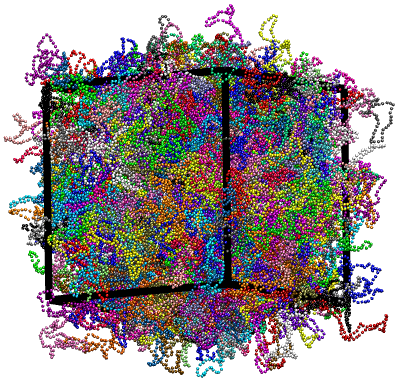}
 \caption{\label{fig:quench} A representative snap-shot of the initial configuration of the system. The system consists of 1000 non-concatenated ring-polymers with 100 monomers per each polymer.}
 \label{fig:initial_conf}
\end{figure}

We use molecular dynamics simulations to study the aging behavior of a system consisting of $N = 1000$ non-concatenated ring polymers, each comprising 100 monomers. A representative snapshot of the system is provided in Fig.~\ref{fig:initial_conf}. The initial configuration is generated by randomly packing the 1000 polymers, followed by equilibration in the NVT ensemble at a temperature of $T = 15$. A Nose-Hoover thermostat is employed to maintain the desired system temperature. After equilibrium is reached, the simulation is continued for an additional $10^8$ time steps with a time step size of $\Delta t = 0.001$. We use this protocol to prepare initial configurations for four different ring polymer stiffness values investigated in this study. After obtaining these equilibrated configurations, we quench the system to a low temperature and run simulations for an additional $3\times10^8$ steps using a larger time step of $\Delta t = 0.005$.

\subsection{Model}

In our system, the monomers in a polymer interact via the finitely extensible non-linear elastic (FENE) potential, 

\begin{equation}
    U_{FENE}(d_{ij})= 
\begin{cases}
    -0.5\kappa R_0^2 ln[1 - (\frac{d_{ij}}{R_0})]&  d_{ij} < R_0\\
    \infty              & d_{ij} \geq R_0.
\end{cases}
\label{eq:fene}
\end{equation}

Here, $d_{ij}$ denotes the distance between two bonded monomers, which must be smaller than $R_0 = 1.5$, the cut-off for the maximum allowable bond length, and $\kappa = 30$ represents the strength of the FENE potential. The FENE potential ensures that polymers cannot be stretched indefinitely. The elasticity of the polymers is modeled using the Kratky–Porod model~\cite{guo2020effects}, given by,

\begin{equation}
    U_{angle}(\theta_i) = K_{\theta}[1 - cos(\theta_i - \pi)].
\label{eq:kratky}
\end{equation}

where, $K_{\theta}$ represents the stiffness of the polymer, and $\theta_i$ is the angle formed at the $i^{th}$ monomer site. In this study, we investigate four distinct stiffness values: $K_{\theta} = 1, 5, 10, 20$, where $K_\theta = 1$ corresponds to flexible polymers, $1 < K_\theta \leq 10$ to semi-flexible polymers, and $K_\theta > 10$ to rigid polymers. Each simulated system is monodisperse, meaning that all polymers within a given system share the same stiffness value. Non-bonded monomer pairs interact via the Lennard-Jones potential,
\begin{equation}
    U_{pair}(r_{ij})= 
\begin{cases}
    4\varepsilon[(\frac{\sigma}{r_{ij}})^{12} - (\frac{\sigma}{r_{ij}})^{6} + \frac{1}{4} ]&  r_{ij}\leq 2^{\frac{1}{6}}\sigma\\
    0              & r_{ij} > 2^{\frac{1}{6}}\sigma,
\end{cases}
\label{eq:lj}
\end{equation}

where $\sigma = 1$ is the monomer diameter, and $\varepsilon$ is the energy scale. For simplicity, we set $\varepsilon = 1$ for our simulations. We truncate the potential so that only repulsive interactions occur at short distances.

\section{\label{sec:results} Results}

\subsection{Identification of thermal glassy regime}
\begin{figure}
\centering
 \includegraphics[width=1.\columnwidth]{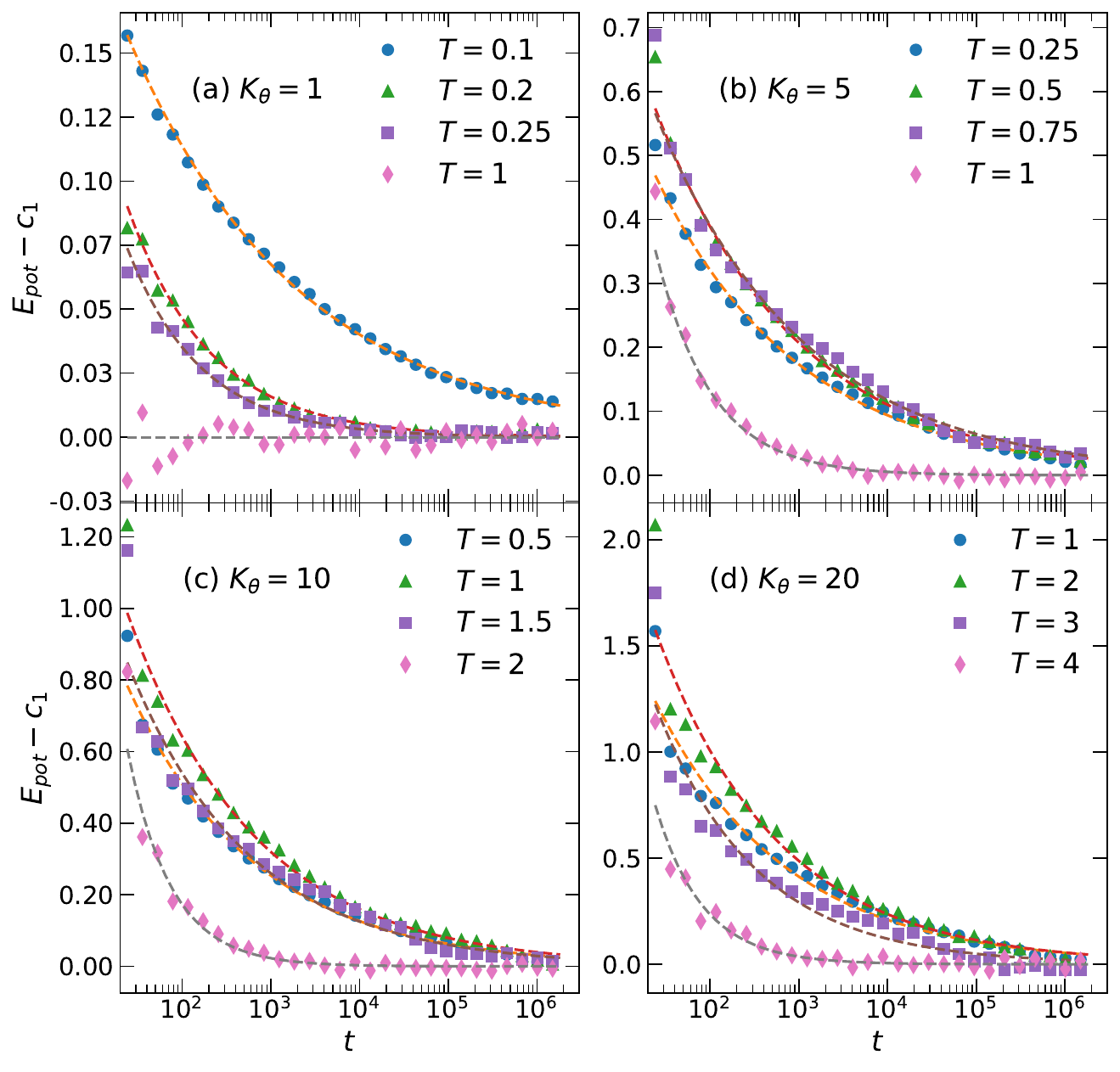}
 \caption{Time variation of potential energy for various polymer stiffness $K_\theta$. The energy is fitted (dashed line) to the algebraic function $E_{pot} = a_1 \tau^{-b_1} + c_1$ and shifted by its intercept ($c_1$). }
 \label{epot_tau_alg}
\end{figure}

We first identify the regime within the phase space of ring stiffness and ambient temperature where the system exhibits glassy dynamics. For each stiffness value ($K_{\theta}$), we quench the system from a high-temperature equilibrium state to various lower target temperatures ($T$). By monitoring the potential energy ($E_{pot}$) over a specified time window following the quench, we ascertain whether the system reaches equilibrium or remains out of equilibrium at these target temperatures~\cite{kob1995testing,berthier2011dynamical,kob2000fluctuations,el2010subdiffusion}. In the supercooled regime, $E_{pot}$ decreases initially and subsequently reaches a plateau, corresponding to its equilibrium value at that temperature. Conversely, in the glassy regime, the potential energy fails to attain a steady-state plateau within accessible simulation timescales. Figure~\ref{epot_tau_alg} presents the time evolution of $E_{pot}$ for different temperatures and polymer stiffness values, illustrating the stiffness-dependent variation in glass transition temperature. For flexible polymers ($K_\theta = 1$), the onset of glassy dynamics occurs around $T \approx 0.1$, where $E_{pot}$ does not stabilize within our observation window. Semi-flexible polymers ($K_\theta = 5$) show glassy behavior at $T \approx 0.75$ but equilibrate at $T = 1$, indicating a slightly higher glass transition temperature compared to flexible polymers. For rigid polymers ($K_\theta = 10, 20$), we progressively observe higher transition temperatures, with clear out-of-equilibrium dynamics emerging at $T \approx 1.5$ for $K_\theta = 10$ and at $T \approx 3$ for $K_\theta = 20$. To quantitatively characterize these relaxation dynamics, we fit the potential energy curves using the algebraic form $E_{pot} = a_1 t^{-b_1} + c_1$ (dashed lines in Fig.\ref{epot_tau_alg}); here, energy curves are shifted by $c_1$ to facilitate direct comparison. The fitting parameter $b_1$ serves as an indicator of structural relaxation timescales, with values between 0.2 and 0.4 representing slow relaxation typical of glassy systems, whereas $b_1 > 0.4$ corresponds to the non-glassy regime. Overall, this analysis delineates the thermal regimes of out-of-equilibrium dynamics, setting the stage for subsequent exploration of aging phenomena\cite{stillinger1988supercooled,el2010subdiffusion}.
\begin{figure}
	\centering
	\includegraphics[width=\columnwidth]{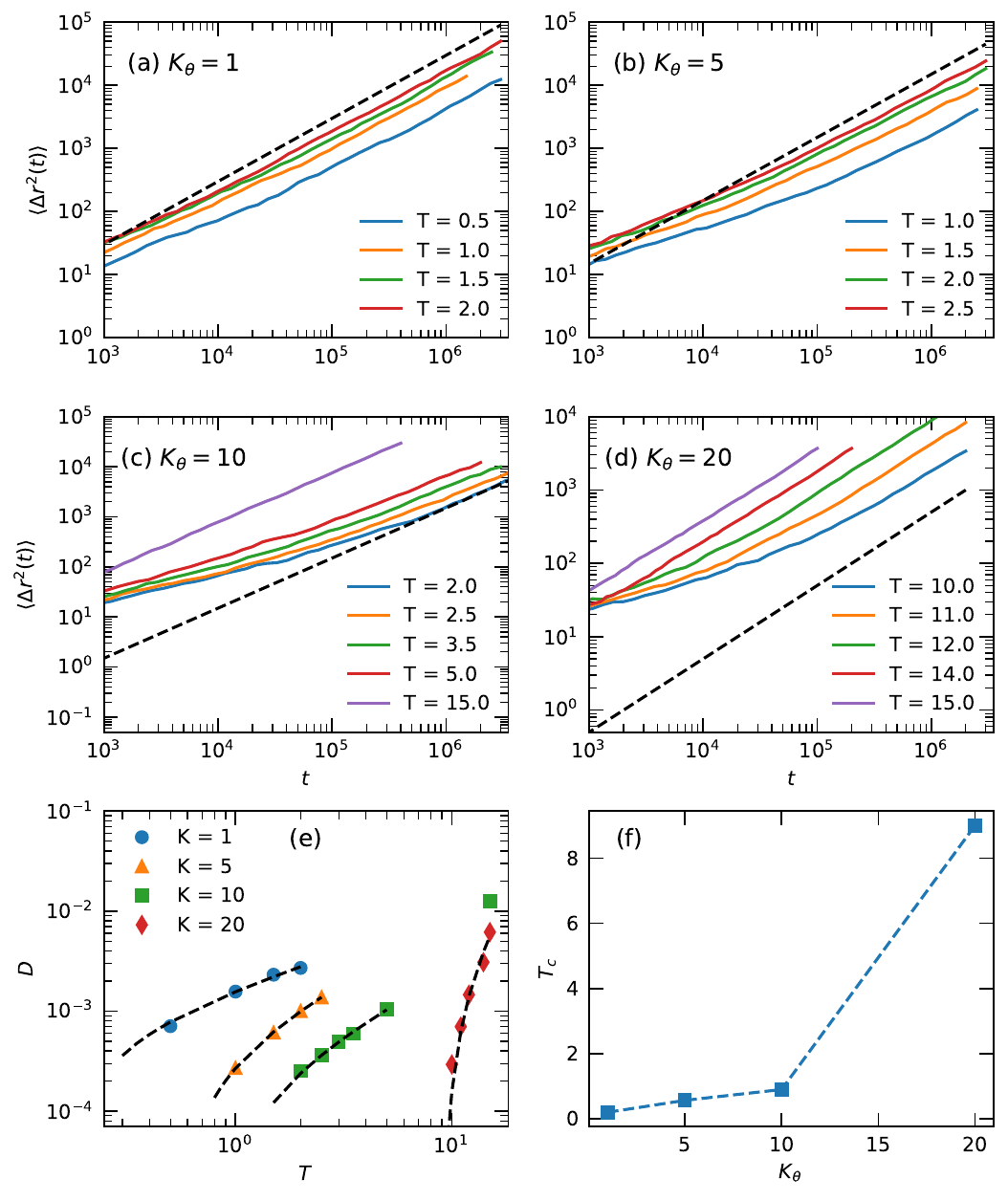}
	\caption{(a-d) Mean square displacement of the center of mass of polymers, $\Delta{r^2}(t)$, for various $K_\theta$ values, where $t$ is the time since thermal quench. (e) Diffusion coefficient $D$ as a function of temperature; the dashed line corresponds to a power-law fit $\sim (T - T_c)^\gamma$ . (f) Estimated $T_c$ for dynamical arrest as a function of polymer stiffness $K_\theta$.}
	\label{rmsd_tw0}
\end{figure}

To obtain a more quantitative estimation of the temperature range at which ring polymer systems of varying stiffness exhibit glassy behavior, we focus on their dynamical properties. Specifically, we identify the temperatures at which diffusive dynamics cease, indicating the onset of dynamical arrest and setting the stage to probe aging phenomena. For this purpose, we measure the mean square displacement (MSD) of the center of mass for the ring polymers, $\Delta r^2(t)$, where $t$ is the time elapsed since the thermal quench from a high-temperature equilibrium state and defined as,
\begin{equation}
\langle\Delta r^2(t)\rangle = \frac{1}{N}\sum_{k=1}^{N} [\vec{R}^k_{cm}(t) - \vec{R}^k_{cm}(0)]^2
\end{equation}
where $\vec{R}^k_{cm}$ is the center-of-mass coordinates of the $k$-the ring.

 Figure~\ref{rmsd_tw0} shows the MSD data across different temperatures and polymer stiffness values, highlighting regimes where diffusion at long times ($\Delta r^2(t) \sim t$) persists. To quantitatively assess the extent of diffusion, we extract the diffusion coefficient ($D$) from these MSD curves and plot its temperature dependence in Fig.~\ref{rmsd_tw0}(e). As anticipated, $D$ decreases systematically as temperature decreases for all polymer stiffnesses. Additionally, at a fixed temperature (e.g., $T = 2$), the diffusion coefficient of flexible polymers exceeds that of stiffer polymers, underscoring the pronounced dependence of diffusion on stiffness ($K_\theta$). Indeed, the diffusion data strongly suggest that dynamical arrest (i.e., $D \rightarrow 0$) occurs at progressively higher temperatures with increasing $K_\theta$.

To estimate the temperature of dynamical arrest ($T_c$) more quantitatively, we perform a power-law fit to the data using the form $D \sim (T - T_c)^\gamma$ (shown as dashed lines in Fig.\ref{rmsd_tw0}(e)). The resulting values of $T_c$ for each stiffness are plotted in Fig.\ref{rmsd_tw0}(f). The onset of non-diffusive dynamics and thus the loss of equilibrium is estimated at temperatures $T_c = 0.197, 0.564, 0.897,$ and $9.0$ for polymer stiffnesses $K_\theta = 1, 5, 10,$ and $20$, respectively. These temperatures serve as reference scales, analogous to the mode-coupling temperature where the diffusion coefficient theoretically vanishes in a power-law manner. Interestingly, the exponent $\gamma$ slightly varies with stiffness, suggesting that increasing $K_\theta$ modifies the nature of the underlying relaxation pathways. Furthermore, the divergence in the $D$ versus $T$ curves indicates a possible stiffness-dependent change in the fragility of these polymeric systems, an aspect that warrants further investigation. The observed increase in $T_c$ with polymer stiffness highlights that stiffer polymers transition into a glassy regime at progressively higher temperatures, reinforcing the role of configurational constraints in hindering relaxation dynamics. In ring polymer systems, topological effects, particularly threading interactions, introduce additional constraints beyond those imposed by conventional polymer entanglement. These constraints elevate $T_c$ beyond what might be anticipated from entanglement alone. This finding aligns with theoretical predictions suggesting that increased configurational restrictions lead to higher energetic barriers for relaxation, shifting the glass transition temperature upward~\cite{privalko1974glass,dudowicz2005glass,hsu2019coarse}. 

In the present study, we observed a notable discrepancy between the estimates of the glassy temperatures as obtained from the potential energy saturation and the $T_c$ estimated from diffusion coefficients extracted from MSD curves, particularly for stiff ring polymers ($K_\theta = 20$). Specifically, the potential energy plateau suggests onset of glassiness around $T \approx 3$, whereas fitting the temperature-dependent diffusion coefficient to a power-law form yields a considerably higher estimate ($T_c \approx 9$). We hypothesize that this difference arises from the distinct length and timescales governing local and global relaxation mechanisms in dynamically heterogeneous, topologically constrained ring polymer melts. Potential energy measurements primarily reflect local structural relaxation at intermediate timescales, capturing short-range polymer segment rearrangements into nearby energetic minima. For stiff ring polymers, local equilibration can occur relatively quickly even at lower temperatures due to minimal dependence on large-scale rearrangements. Conversely, MSD measurements probe global structural mobility and translational diffusion, strongly influenced by long-range topological constraints such as threading. Threading interactions impose severe constraints on global rearrangements, drastically suppressing diffusion, thus potentially shifting diffusion-based $T_c$ estimates to higher temperatures. The distinct estimates of $T_c$ therefore highlight the possibility of multiple characteristic temperatures in topologically constrained polymeric glasses—one associated with local structural equilibration, and another related to global dynamic arrest.

\subsection{Aging dynamics}

\begin{figure}
	\centering
	\includegraphics[width=\columnwidth]{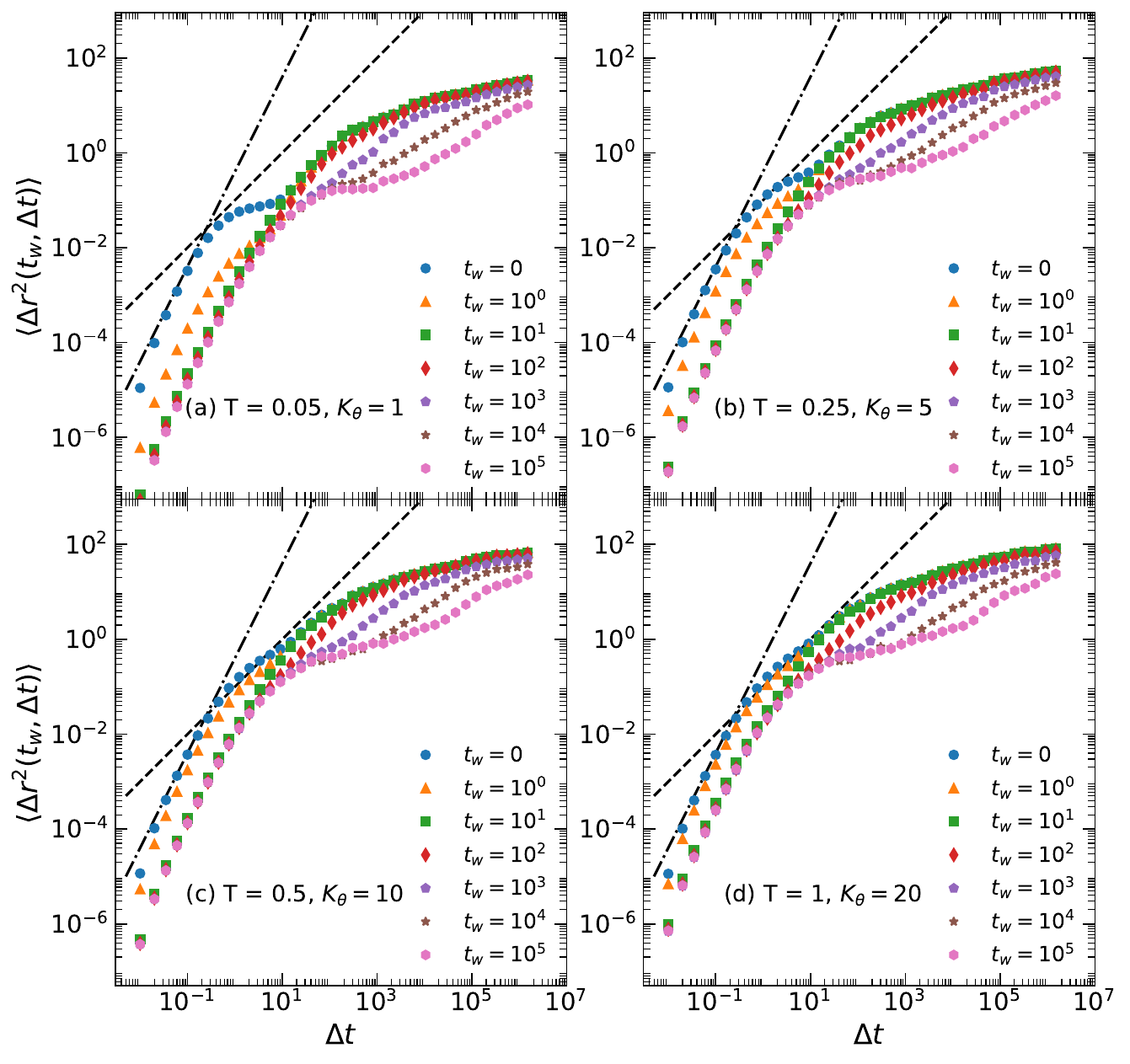}
	\caption{ Mean square displacement of the center of mass of polymers, $\Delta{r^2}(t_w,\Delta{t})$, as a function of waiting time for different stiffness values and temperatures. }
	\label{fig:msd_wait}
\end{figure}

\begin{figure}
	\centering
	\includegraphics[width=\columnwidth]{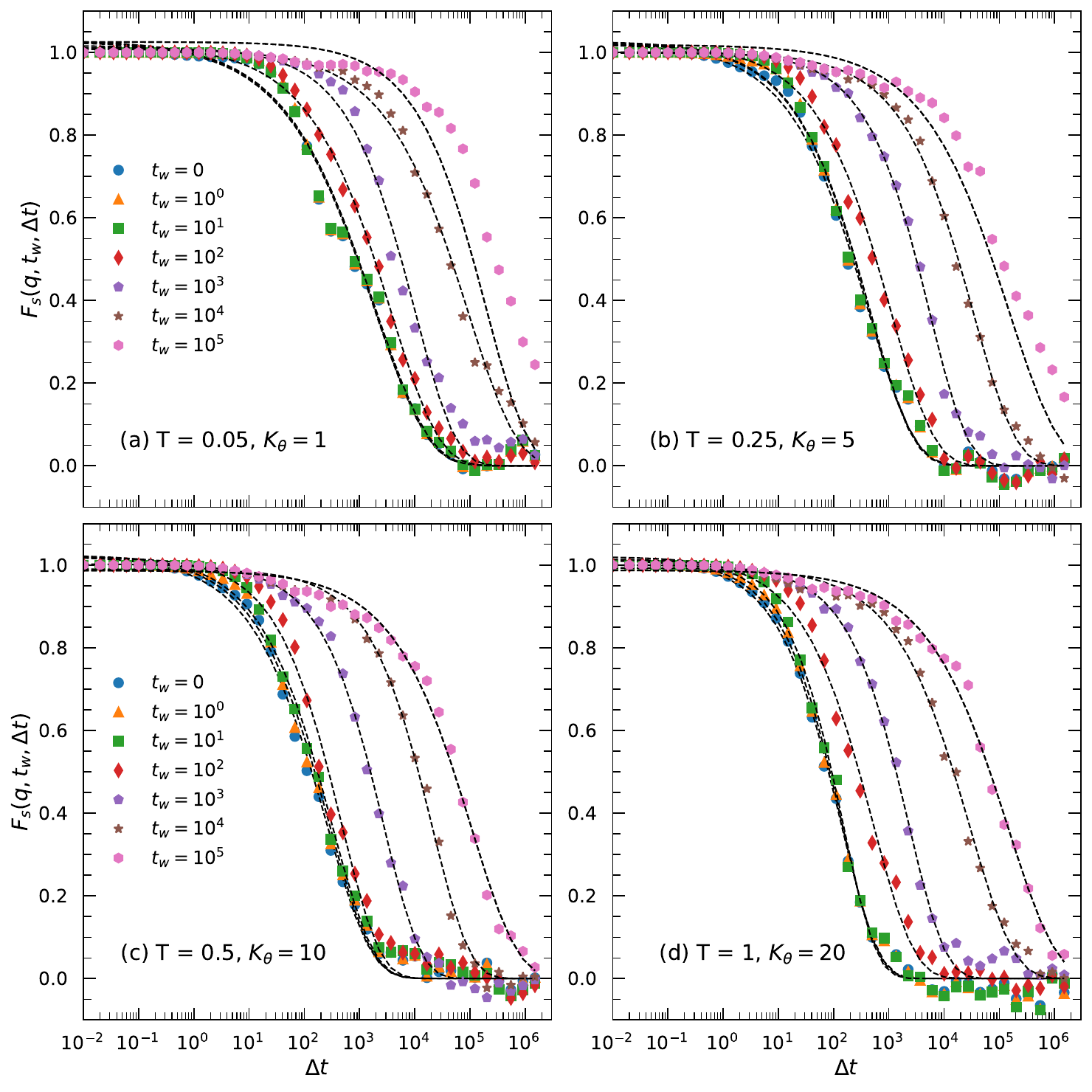}
	\caption{Waiting time dependency of the self-intermediate scattering function (ISF), $F_s(q, t_w, t)$ for $K_\theta = 1, 5, 10, 20$, computed for $q=1$. The dashed lines correspond to fits with stretched exponential functions.}
	\label{fig:wt_fqt_allK}
\end{figure}

In equilibrated systems, temporal correlation functions depend solely on the time difference between measurements, reflecting time-translational invariance. However, in glassy systems, equilibration becomes increasingly challenging within accessible simulation or experimental timescales as the ambient temperature approaches and eventually drops below the characteristic temperature ($T_c$). This arises because relaxation times rapidly increase with decreasing temperature, causing the system to fall progressively out of equilibrium. Consequently, correlation functions in such systems depend not only on the time difference but also explicitly on the initial measurement time, known as the waiting time ($t_w$), making them two-time quantities~\cite{kob2000fluctuations}. This $t_w$ dependence is a hallmark of aging behavior.

To explore aging dynamics in our system, we analyze the waiting-time-dependent mean square displacement (MSD) of the polymer center-of-mass (COM) and the intermediate scattering function (ISF). We particularly investigate how polymer stiffness ($K_\theta$) and temperature ($T$) influence dynamical slowdown, sub-diffusive behavior, and aging phenomena, thereby elucidating distinctive features of aging in ring polymer systems compared to conventional polymer and colloidal glasses. Figure~\ref{fig:msd_wait} presents the waiting-time-dependent MSD for ring polymers following instantaneous quenches from a high initial temperature to target temperatures below their respective $T_c$ values (indicated in each sub-plot) for each stiffness $K_{\theta}$. At short timescales ($\Delta{t}\equiv{t-t_w} < 10$), the MSD curves exhibit mild waiting-time dependence at small $t_w$ due to transient memory effects from the high-temperature initial state. However, as the system settles at the lower ambient temperature, early ballistic dynamics become independent of $t_w$. At intermediate to longer timescales, the MSD curves consistently display sub-diffusive behavior and prominent dependence on waiting time. This dependence grows stronger at higher $t_w$, reflecting a gradual dynamical slowdown as the system ages. Aging signatures appear clearly as plateaus in $\Delta{r^2}(t_w,\Delta{t})$ at larger waiting times, indicative of particle caging effects arising from neighboring rings. These aging characteristics observed in ring polymer MSD align qualitatively with other soft glassy systems, such as colloidal hard spheres~\cite{el2010subdiffusion}. However, the ring polymers' unique topology, notably threading interactions, imposes additional constraints influencing their aging dynamics, which we examine in subsequent sections.
\begin{figure}
\centering
 \includegraphics[width=\columnwidth]{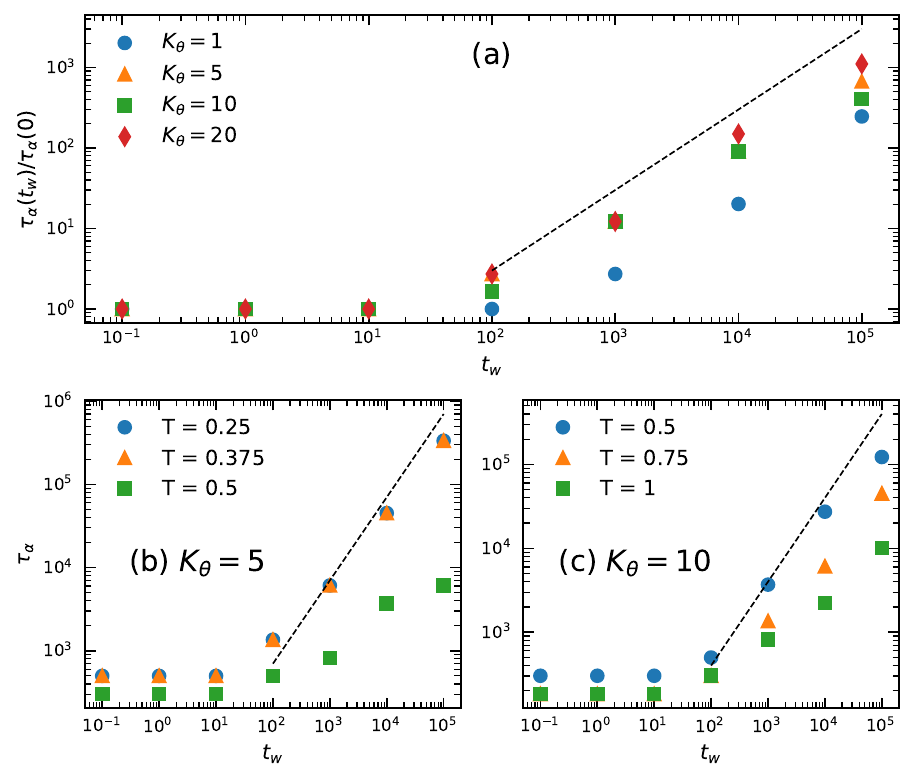}
 \caption{(a) Relaxation time $\tau_\alpha$ as a function of waiting time $t_w$ for state points shown in Fig.\ref{fig:wt_fqt_allK}, with the timescales rescaled by $\tau_\alpha(0)$. (b) and (c) highlight how $\tau_\alpha(t_w)$ varies with temperature, as shown for $K_\theta = 5$ and $K_\theta = 10$, respectively. The dashed line in each sub-plot corresponds to simple aging scenario of $\tau_\alpha \sim t_w$.}
 \label{fig:tauq}
\end{figure}

The age dependent self-intermediate scattering function (ISF), $F_s(q, t_w, t)$, quantifies the temporal correlation of density fluctuations and is widely used to characterize relaxation dynamics in glass-forming liquids and is defined as,
\begin{equation}
F_s(q,t_w, t) = \frac{1}{N}\sum_{k=1}^{N} e^{i\vec{q}.[R^k_{cm}(t+t_w) - R^k_{cm}(t_w)]}
\end{equation}
where the summation is over all ring polymers and $R^k_{cm}$ is the center of mass of $k$th ring polymer. Unlike the simple exponential decay observed in liquids, supercooled liquids display a slow, non-exponential decay of the ISF. This occurs due to activated hopping processes between multiple minima in a rugged potential energy landscape, becoming increasingly prominent as the glass transition temperature is approached~\cite{kob2000fluctuations,berthier2011dynamical}. Like the MSD discussed above, the ISF should also exhibit aging behavior, which we illustrate in Fig.\ref{fig:wt_fqt_allK} through the waiting-time-dependent correlation function $F_s(q, t_w, t)$, measured for $q = 1$, corresponding to length scales of approximately six monomer diameters—larger than the typical cage size evident from MSD plateaus (Fig.\ref{fig:msd_wait}). At long timescales, the ISF decays increasingly slowly with larger $t_w$, indicating a progressive slowdown in structural relaxation dynamics as the system ages. As shown in Fig.\ref{fig:wt_fqt_allK}, it is possible to find reasonable fits of the data with stretched exponential functions.

The aging-dependent relaxation is quantified via a relaxation time, $\tau_\alpha(t_w)$, defined as the time at which $F_s(q, t_w, \tau_\alpha) = 1/e$. Figure~\ref{fig:tauq} shows the waiting-time dependence of $\tau_\alpha$ for different polymer stiffness values, $K_\theta$. For reference, we include a dashed line corresponding to a simple aging scenario, $\tau_\alpha \sim t_w$. Figure~\ref{fig:tauq}(a) compares examples across various stiffnesses at temperatures discussed previously, with $\tau_\alpha(t_w)$ rescaled by $\tau_\alpha(0)$ for comparison. For small waiting times, $\tau_\alpha$ remains nearly constant, consistent with transient relaxation dominated by initial temperature effects (as also visible in MSD data, Fig.~\ref{fig:msd_wait}). However, for $t_w > 10^2$, $\tau_\alpha$ significantly increases with aging, and stiffer polymers display a more pronounced rise compared to flexible ones ($K_\theta = 1, 5$), underscoring stiffness-enhanced dynamical constraints.Figure~\ref{fig:tauq}(b-c) further examines temperature effects on aging dynamics for intermediate and rigid stiffness values, $K_\theta=5,10$. For semi-flexible polymers ($K_\theta = 5$, Fig.~\ref{fig:tauq}(b)), at lower temperatures ($T = 0.25$), the increase in relaxation time closely follows the simple aging scenario at intermediate $t_w$, with slight sublinearity at longer ages. Conversely, at higher temperatures closer to $T_c$ ($T = 0.5$), $\tau_\alpha$ grows more slowly and tends toward a plateau at larger $t_w$, indicating the system remains close to equilibrium. Rigid polymers ($K_\theta = 10$, Fig.\ref{fig:tauq}(c)) exhibit a similar trend.

The observed waiting-time dependence of MSD and ISF is qualitatively consistent with studies of aging in colloidal glasses and dense polymer systems~\cite{kob2000fluctuations,el2010subdiffusion}. However, the unique topology and stiffness of ring polymers introduce additional constraints, leading to distinct aging signatures, including sublinear growth of $\tau_\alpha$ with $t_w$. Increased stiffness amplifies excluded volume effects, restricting configurational rearrangements and intensifying aging behavior. These findings align with studies of dense polymer melts, where stiffness delays relaxation and enhances dynamical heterogeneity~\cite{rubinstein2003polymer}. The pronounced dependence of $\tau_\alpha$ on $t_w$ and $K_\theta$ highlights the crucial roles played by both stiffness and waiting time in dictating the aging dynamics of glassy ring polymer systems. Moreover, the absence of free ends and presence of threading in ring polymers impose unique topological constraints that further slow structural relaxation, distinctly differentiating their aging dynamics from linear polymeric systems. We examine these topological effects, particularly threading, and their implications on aging dynamics in the next section.

\subsection{Threading Analysis}

Ring polymers, unlike linear polymers, are topologically constrained, making structural rearrangements difficult, particularly at low temperatures. The absence of free ends prevents ring polymers from undergoing reptation, the primary relaxation mechanism in linear polymer melts. Instead, their unique topology allows for threading, a phenomenon where one ring polymer passes through the contour of another, introducing additional entanglement constraints~\cite{roy2022effect}. Formally, threading is defined by the intersection of bond vectors (monomer–monomer bonds) of one ring polymer with the triangular planes forming the minimal surface of another ring polymer. Specifically, if the bond vectors of ring A intersect with the minimal surface of ring B, ring A is said to thread into ring B. To quantify threading, we define $N_{th}$ as the number of bonds of ring A located between two successive intersections. To simplify calculations, we divide each ring into an odd side and an even side and count the bonds on the odd side ($N_{th}^{odd}$). Given the absence of free ends, the total number of bonds satisfies the relation $N_{th}^{odd} + N_{th}^{even} = N_{ring}$, where $N_{ring}$ is the total number of bonds in a ring. Finally, we define the degree of threading as:

\begin{equation} 
\Delta(t) = 1 - \left\lvert \frac{2\sum\limits_{i=odd} N_{th}^i}{N_{ring}} - 1 \right\rvert. 
\label{eq_threading} 
\end{equation}

Equation~\ref{eq_threading} provides a quantitative measure of threading events in the system. A value of $\Delta(t) = 0$ corresponds to no threading, i.e., when $\sum\limits_{i=odd} N_{th}^i = 0$, whereas a maximum value of $\Delta(t) = 1$ occurs when a ring is perfectly bisected by another, such that $\sum\limits_{i=odd} N_{th}^i = N_{ring}/2$. Applying Equation~\ref{eq_threading}, we compute the threading degree for each ring polymer and obtain an average over all rings to quantify the system-wide degree of threading over time. We first analyze how the extent of threading evolves as the system is quenched from a high-temperature liquid state into a low-temperature topological glassy state. Since a topological glass is characterized by a high degree of threading, it is expected that the initially equilibrated high-temperature liquid will exhibit significantly lower threading. 


\begin{figure}
\centering
 \includegraphics[width=0.9\columnwidth]{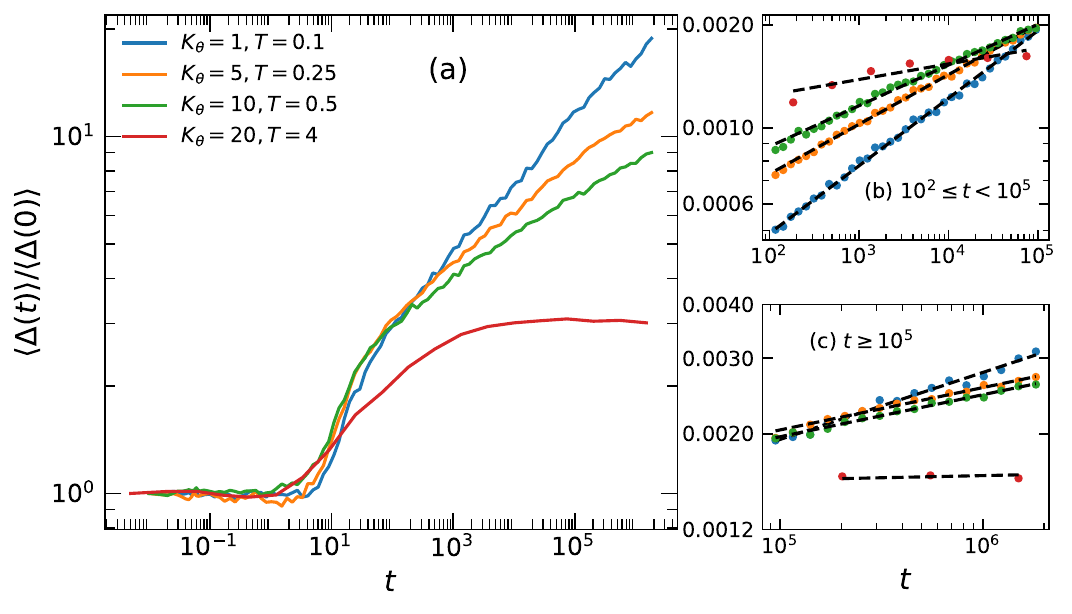}
 \includegraphics[width=0.9\columnwidth]{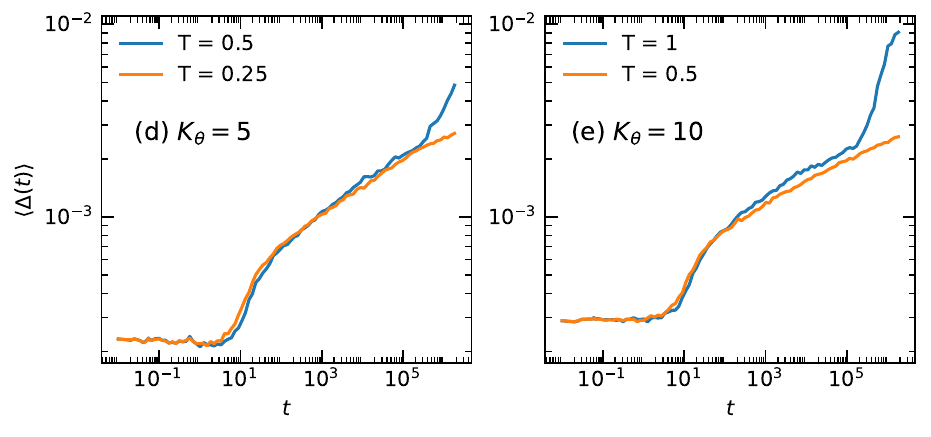}
  \includegraphics[width=0.9\columnwidth]{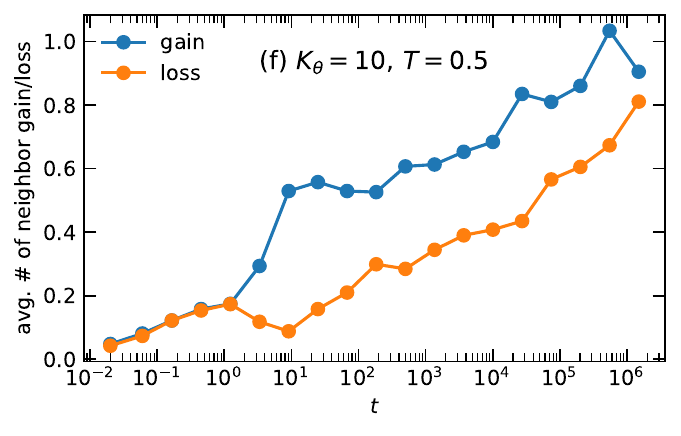}
 \caption{(a) Time evolution of the degree of threading relative to initial state, $\Delta(t)/\Delta(0)$, for various stiffness values $K_\theta = 1, 5, 10, 20$ at their respective glassy state points discussed in Fig.\ref{fig:wt_fqt_allK}. (d, e) Time evolution of $\Delta(t)$ at two different temperatures, for  $K_\theta = 5, 10$ respectively.  (f) Number of threading neighbor gain/loss events over time, measured at the state point $K_\theta = 10, T = 0.5$.}
 \label{fig:threading}
\end{figure}

Indeed, this is observed in Figure~\ref{fig:threading}(a) which shows the time evolution of the average degree of threading $\Delta(t)$ for ring polymers with varying stiffness values ($K_\theta = 1, 5, 10, 20$) following a quench to temperatures below their respective $T_c$. At short times ($t < 10$), threading remains nearly constant across all stiffnesses, as expected due to the persistence of high-temperature initial conditions and the dominance of local fluctuations. For $K_\theta = 1$, $5$, and $10$, threading increases gradually and with similar overall trends. Despite differences in stiffness, these systems display comparable ability to explore topological configurations that enable threading events. This suggests that up to moderate stiffness, polymers retain sufficient conformational flexibility and configurational freedom for threading to develop progressively over time, without severe hindrance. In contrast, the $K_\theta = 20$ system exhibits markedly different behavior. While initial growth is visible, threading saturates rapidly and remains flat at long timescales, indicating strong dynamical arrest. The high stiffness in this regime likely imposes significant steric and topological constraints, suppressing large-scale rearrangements required for further interpenetration. This stagnation in $\Delta(t)$ growth reflects a glassy state dominated by persistent threadings and lack of mobility. These results highlight that while threading dynamics are moderately sensitive to stiffness in the flexible-to-semi-flexible regime, a qualitative shift occurs at high stiffness: configurational constraints dominate, and threading becomes dynamically frozen. To further quantify the long-time dynamics of threading, the data for $t \geq 10^5$ are fitted to a power-law relationship, $\Delta(t) \sim t^\eta$. Figure~\ref{fig:threading}(b) shows the fits at early time, with the exponent $\eta$ capturing the stiffness dependence of long-time threading behavior. The values of $\eta$ range from 0.04 to 0.2 and these results highlight the distinct dynamic regimes introduced by polymer stiffness. Figure~\ref{fig:threading}(c) shows the fits at long time, with all stiffness values, except $K_\theta=20$ show similar trend, clearly suggesting that the highest stiffness value considered here shows a distinct behavior.

In Figure~\ref{fig:threading}(d-e), we analyze how temperature affects threading growth for polymers with stiffness $K_\theta = 5$ and $K_\theta = 10$, by comparing threading dynamics at two temperatures: one significantly below $T_c$ and another near $T_c$. Initially, threading growth is similar at both temperatures; however, at later times, distinct deviations emerge. At higher temperatures near $T_c$, threading growth shows a secondary rapid increase at long times. This suggests that enhanced thermal fluctuations near $T_c$ facilitate additional structural rearrangements, allowing further threading events, a mechanism suppressed at lower temperatures deep in the glassy regime.

Figure~\ref{fig:threading}(f) provides further microscopic insights by tracking the number of gain and loss events associated with threading over time, specifically for polymers with $K_\theta = 10$ quenched to a glassy state at $T = 0.5$. The gain curve increases more steeply compared to the loss curve, indicating that new threading events continue to form during the system’s evolution and, once established, threading constraints are not easily undone. At short times ($t < 10$), both gain and loss rates remain low, consistent with limited large-scale rearrangements. With increasing time, the rate of threading events accelerates, signaling the onset of structural reorganization. However, threading loss events remain comparatively infrequent, highlighting the persistent nature of topological constraints formed during aging. The sustained increase in threading reinforces the notion that ring polymers experience durable topological constraints, particularly at low temperatures, leading to significantly slowed dynamics and characteristic glassy rigidity. To quantify aging effects on threading dynamics, we calculate the average threading duration between pairs of rings. We define the threading matrix $P_{ij}(t)$ at time $t$ as follows:
\begin{equation}
    P_{ij} (t)= 
    \begin{cases}
        1, & \text{if } i^{th} \text{ ring} \text{ is threaded to } j^{th}\\
        0, & \text{otherwise.}
    \end{cases}
    \label{eq:threading}
\end{equation}
Here, $P_{ii} = 0$, i.e., self-threading events are excluded from this calculation. Also, when the $i^{th}$ polymer is threaded to the $j^{th}$ polymer, $P_{ij} = 1$, but the converse is not necessarily true, i.e., $P_{ji} \neq 1$. At any point in time, $P_{ij}(t)$ is an $N \times N$ matrix, where N is the number of polymers. Using this matrix, we define the correlation function \cite{michieletto2016topologically},

\begin{equation}
\phi_{p}(t_w, t) = \langle \frac{1}{N} \sum_{j = 1}^{N} P_{ij}(t+t_w).P_{ij}(t+t_w-\delta t)...P_{ij}(t_w)\rangle.
\end{equation}

where $\langle \dots \rangle$ denotes the average over all rings. When two rings unthread, $P_{ij} (t)=1\rightarrow{0}$, and therefore the correlation function $\phi_{p}(t_w, t)$ tracks such unthreading events, as a function of waiting time $t_w$, thereby measuring threading persistence. 

\begin{figure}
\centering
 \includegraphics[width=0.95\columnwidth]{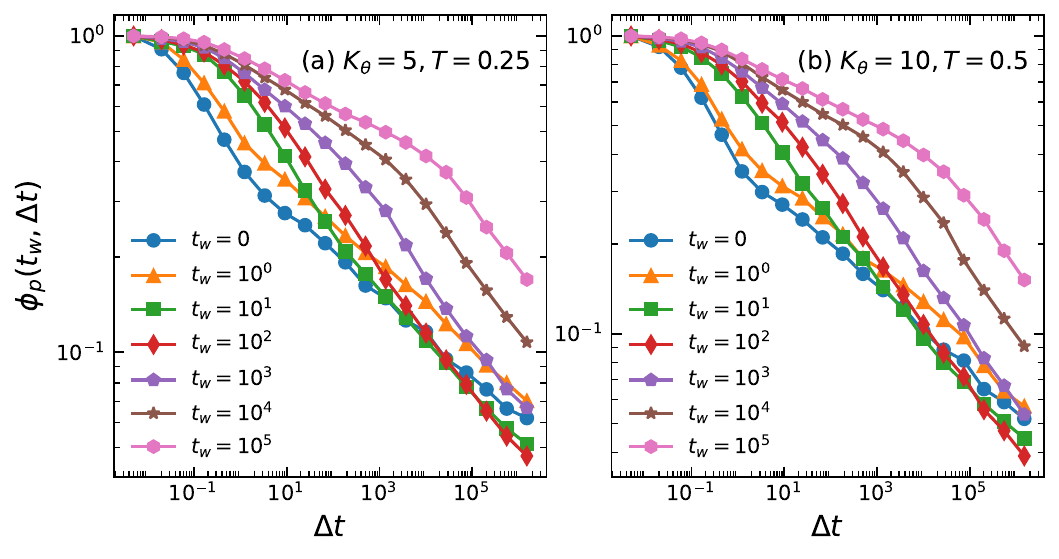}
\includegraphics[width=0.95\columnwidth]{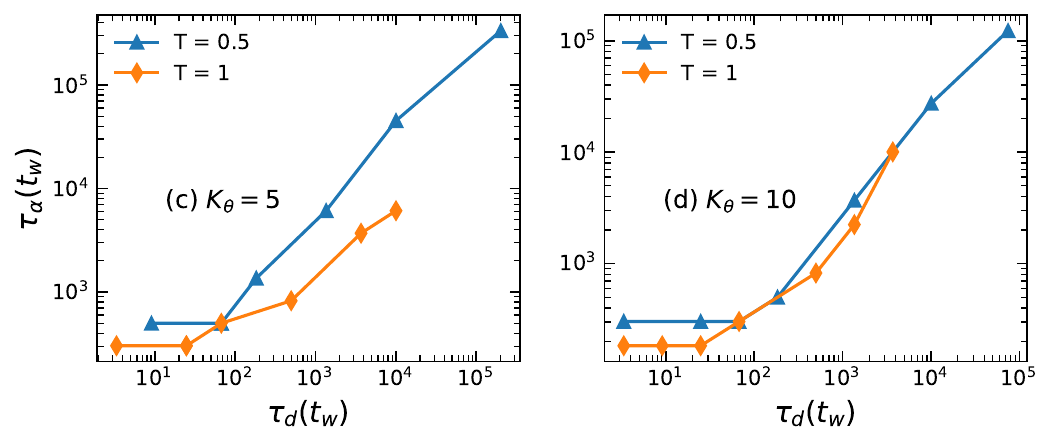}
 \caption{Age dependence of the overlap function $\phi_{p}(t_w, \Delta t)$ measured at the state points: (a) $K_{\theta} = 5$, $T = 0.25$ and (b) $K_{\theta} = 10$, $T = 0.5$.  (c) Variation of the characteristic discontinuity time, $\tau_d$, extracted from $\phi_{p}(t_w, \Delta t)$, with age $t_w$, shown for $K_{\theta} = 10$, at temperatures of $T=0.5, 1.0$. (d) Correlation of structural relaxation time, $\tau_\alpha$, with threading discontinuity time, $\tau_d$, as obtained via their respective age dependence and computed for the state points shown in (c).}
 \label{fig:overlap}

\end{figure}

Figure~\ref{fig:overlap}(a-b) shows the waiting time dependence of $\phi_p(t_w,t)$ for ring polymers with stiffness values $K_{\theta} = 5$ and $K_{\theta} = 10$. At short times ($t < 10$), $\phi_p(t_w,t) \approx 1$ across all waiting times, reflecting the ballistic regime in which monomers remain highly correlated with their initial positions. The initial sharp drop for smaller $t_w$ values is attributed to transient effects following the high-temperature quench. As time progresses, the decay of $\phi_p(t_w,t)$ slows significantly for larger $t_w$, demonstrating progressive dynamical slowdown, which is characteristic of aging in glassy systems. Thus, threading decorrelation effectively captures signatures of aging dynamics. We quantify this threading decorrelation using the characteristic discontinuity timescale ($\tau_d$), defined as the time at which $\phi_p(\tau_d) = 0.3$, corresponding to 70\% decorrelation of initially threaded rings. Figure~\ref{fig:overlap}(c) shows that $\tau_d$ increases systematically with waiting time, $t_w$, indicating that the unthreading process slows as the system ages. Thermal effects are also evident; at higher temperatures, $\tau_d$ is smaller, and the unthreading becomes notably faster, suggesting that reduced thermal fluctuations amplify the role of topological constraints in glassy dynamics. Finally, we examine the relationship between threading persistence and single-ring relaxation dynamics by correlating the threading timescale ($\tau_d$) with the structural relaxation time ($\tau_\alpha$) derived from the ISF. Figure~\ref{fig:overlap}(d) clearly demonstrates a positive correlation between $\tau_d$ and $\tau_\alpha$, underscoring threading as a critical factor governing long-term relaxation in glassy ring polymer systems. These findings emphasize the unique aging behavior of ring polymers, distinctly influenced by topological constraints compared to conventional polymeric or colloidal glasses.

\section{\label{sec:conclusions} Summary and Discussion}

In this study, we investigated the aging dynamics and glass transition behavior of dense, topologically constrained ring polymer systems using large-scale molecular dynamics simulations. First, by monitoring the long-time diffusion of the center of mass of the rings in the supercooled regime, we established that the temperature ($T_c$) at which the system becomes non-ergodic increases steadily with polymer stiffness ($K_\theta$), highlighting rigidity’s role in enhancing configurational constraints and increasing the energy barriers for relaxation. This behavior aligns with prior findings in linear polymer systems and colloidal glasses, reinforcing the concept that increased stiffness amplifies structural resistance to thermal fluctuations, thus shifting the glass transition to higher temperatures.

Below $T_c$, hallmark characteristics of aging were evident. Analysis of the mean square displacement (MSD) and intermediate scattering function (ISF) revealed pronounced non-equilibrium dynamics with the aging dynamics manifested clearly through the waiting-time ($t_w$) dependence of these correlation functions.  The relaxation time ($\tau_\alpha$) followed a sublinear power-law relationship ($\tau_\alpha \sim t_w^b$), with the exponent $b$ ranging from 0.8 to 0.93, suggesting a similarity with the simple aging scenario  observed in colloidal hard spheres.

The threading behavior, a phenomenon specific to ring polymers, provided additional insights into the interplay between stiffness, topology, and glassy dynamics. Our findings indicated that the degree of threading, quantified by the metric $\Delta(t)$, evolved distinctly across stiffness regimes and time scales. At early times ($t < 10$), threading events remained minimal across stiffness values, reflecting transient high-temperature effects and local rearrangements. As the system entered the glassy regime, significant stiffness-dependent threading behaviors emerged. Flexible polymers ($K_\theta = 1$) showed steady but moderate threading increases, consistent with their compact conformations limiting ring penetration. Semi-flexible polymers ($K_\theta = 5$) demonstrated optimal threading growth, balancing configurational flexibility and excluded volume constraints. Moderately rigid polymers ($K_\theta = 10$) exhibited initially slow threading dynamics, but at longer timescales ($t > 10^5$), threading increased markedly due to delayed cooperative rearrangements. Highly rigid polymers ($K_\theta = 20$) rapidly saturated in threading events, dominated by steric hindrance and dynamic arrest. Long-time threading followed a power-law scaling, $\Delta(t) \sim t^\gamma$, with exponent $\gamma$ systematically varying with stiffness. Polymers of intermediate stiffness ($K_\theta = 5, 10$) exhibited higher $\gamma$ values, indicative of greater structural rearrangements promoting threading, whereas flexible and highly rigid polymers had lower $\gamma$, reflecting constraints from compactness and steric hindrance. Analysis of threading gain and loss further revealed that, as aging progressed, threading interactions became increasingly persistent; the gain of new threadings slowed, while existing interactions became long-lived, suggesting threading’s stabilizing role in the glassy state.

To further probe relaxation behavior, we analyzed the overlap function $\phi_p(t_w,t)$, quantifying local structural persistence. The decay of $\phi_p(t_w,t)$ slowed with increasing $t_w$, highlighting growing resistance to structural rearrangements with aging. This aging effect was pronounced for stiff polymers ($K_\theta = 10$), where relaxation dynamics slowed significantly at large $t_w$. The characteristic relaxation time ($\tau_{\alpha}$) extracted from $\phi_p(t_w,t)$ displayed a similar power-law dependence on waiting time ($\tau_{\alpha} \sim t_w^b$), with exponents consistent with MSD and ISF analyses. The overlap function further underscored polymer stiffness’s influence on structural persistence; flexible polymers relaxed faster, whereas stiff polymers retained structural correlations significantly longer, reinforcing stiffness-induced topological constraints and pronounced aging. The observed dependencies of MSD, ISF, threading, and overlap functions on polymer stiffness and waiting time provide valuable insights into ring polymers’ distinctive aging dynamics. Unlike conventional colloidal systems, the topological constraints in ring polymers—particularly threading—significantly amplify dynamical slowdown and aging effects with increasing stiffness. Moreover, deviations from diffusive motion and strong waiting-time dependence confirm the inherently glassy nature of these systems below $T_c$. Our findings emphasize the critical interplay between polymer stiffness, topology, and aging dynamics, shedding new light on the fundamental behavior of glassy ring polymer systems. Collectively, our results establish polymer stiffness as a key parameter controlling glass formation in ring polymer systems, distinguishing them from conventional polymeric and colloidal glasses. The interplay between increased stiffness and threading-induced topological frustration yields unique dynamical slowdowns, providing novel insights into how polymer architecture influences the glass transition process.

\begin{acknowledgments}
We thank the HPC facility (Nandadevi, Kamet) at The Institute of Mathematical Sciences for computing time. P.C. and S.V. also acknowledge support via the subproject on the Modeling of Soft Materials within the IMSC Apex project, funded by Department of Atomic Energy.
\end{acknowledgments}

\bibliography{ref}
\end{document}